\documentclass[a4paper,11pt]{article}
\pdfoutput=1 

\usepackage{jheppub} 
\usepackage{graphicx}
\usepackage{amsmath}
\usepackage{slashed}
\usepackage{amssymb}
\usepackage[mathscr]{euscript}
\usepackage{enumerate}
\usepackage{hyperref}

\hypersetup{
	colorlinks=true,
	linktoc=page,
	linkcolor=blue,
	citecolor=blue,
	urlcolor=blue
	}

\newcommand{\J}[0]{ {\mathcal J} }
\newcommand{\grad}[0]{ {\vec{\partial} } }

\usepackage[T1]{fontenc} 

\title{Einstein-Maxwell Dirichlet walls, negative kinetic energies, and the adiabatic approximation for extreme black holes}

\author[a]{Tom\'as Andrade}
\author[b]{William R. Kelly}
\author[b]{Donald Marolf}

\affiliation[a]{Rudolf Peierls Centre for Theoretical Physics \\ University of Oxford, 1 Keble Road, Oxford OX1 3NP, United Kingdom}
\affiliation[b]{University of California at Santa Barbara \\ Santa Barbara, CA 93106, USA}

\emailAdd{tomas.andrade@physics.ox.ac.uk}
\emailAdd{wkelly@physics.ucsb.edu}
\emailAdd{marolf@physics.ucsb.edu}

\abstract{The gravitational Dirichlet problem -- in which the induced metric is fixed on boundaries at finite distance from the bulk -- is related to simple notions of UV cutoffs in gauge/gravity duality and appears in discussions relating the low-energy behavior of gravity to fluid dynamics.  We study the Einstein-Maxwell version of this problem, in which the induced Maxwell potential on the wall is also fixed. For flat walls in otherwise-asymptotically-flat spacetimes, we identify a moduli space of Majumdar-Papapetrou-like static solutions parametrized by the location of an extreme black hole relative to the wall.  Such solutions may be described as balancing gravitational repulsion from a negative-mass image-source against electrostatic attraction to an oppositely-signed image charge.  Standard techniques for handling divergences yield a moduli space metric with an eigenvalue that becomes negative near the wall, indicating a region of negative kinetic energy and suggesting that the Hamiltonian may be unbounded below. One may also surround the black hole with an additional (roughly spherical) Dirichlet wall to impose a regulator whose physics is more clear. Negative kinetic energies remain, though new terms do appear in the moduli space metric.  The regulator-dependence indicates that the adiabatic approximation may be ill-defined for classical extreme black holes with Dirichlet walls.
}

\begin{document}
\maketitle
\flushbottom

\section{Introduction}

Boundary conditions are an important part of any physical problem.  For gravitational problems they are often imposed by restricting the falloff of the metric at infinity.  Such asymptotic boundary conditions are appropriate for isolated physical systems.  But there are also applications in which it is interesting to study the dynamics of a finite region of spacetime.

A well known example is the study of thermodynamics for ``gravity in a box'' (see e.g.~\cite{Hawking:1976de,Brown:1994su}).  In this regard it is interesting to note that simple black-hole-in-box calculations based on stationary black holes indicate that such regulated systems have bounded entropy and energy, as one would expect for field theories in finite volume with a UV cut-off, although the temperature diverges as the black hole horizon approaches the wall of the box.  The apparent bound on area follows from the requirement that the black hole fit inside a box of fixed size, while the apparent bound on energy is the statement that Schwarzschild black holes have finite energy in the limit where their size approaches that of the box \cite{Brown:1992br}; see however section 5 of \cite{Andrade:2015gja} for further comments.

Taking the limit where the box walls approach the black hole horizon has been argued to lead to a controlled version of the black hole membrane paradigm~\cite{Damour:1978cg,thorne1986black}.  In particular, Bredberg et al.~\cite{Bredberg:2010ky,Bredberg:2011jq} found in this limit that gravitational dynamics reduced to that of a fluid living on the cutoff surface (see also~\cite{Brattan:2011my,Tian:2014goa}).

In addition, the walls of our box can also be used to model the confining potential of AdS space.  Indeed, in the AdS/CFT context it is common practice \cite{Maldacena:1997re,Witten:1998qj,Susskind:1998dq,Henningson:1998gx,Peet:1998wn,Balasubramanian:1999re} to regulate AdS calculations by restricting to a finite volume $V$ and fixing the metric on $\partial V$, thus imposing what from the CFT perspective is a surprising Lorentz-invariant UV regulator.  See also \cite{Bredberg:2010ky,Heemskerk:2010hk,Faulkner:2010jy} for related discussions of Wilsonian renormalization.

The above discussions highlight the need for a more complete understanding of gravitational dynamics in the presence of such boundaries.   We take preliminary steps in this direction below by investigating the stability of Einstein-Maxwell theory near a Dirichlet wall.  We focus on the case where the wall metric is Minkowski and the induced Maxwell potential vanishes on the wall.  Far from the wall we take the spacetime to be asymptotically flat.  For definiteness we work in 3+1 dimensions.

\begin{figure}
\centering
\includegraphics[width=0.25 \textwidth]{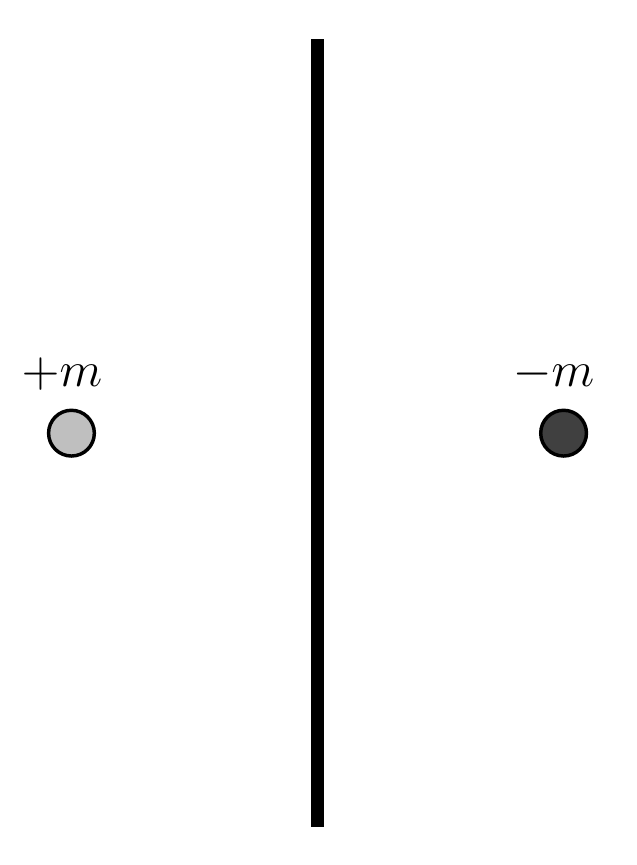}
\caption{At the linearized level the Dirichlet wall will repel uncharged black holes, as can be seen using the method of images.  However, if these black holes are extremal and oppositely charged, the electrostatic attraction precisely cancels the gravitational repulsion.}
\label{fig:Image}
\end{figure}

We will use the term {\it Dirichlet wall} to denote any finite surface on which the induced metric is fixed. It remains an open problem whether the gravitational initial value problem remains well-defined at the non-linear level with such boundary conditions (though see \cite{Szilagyi:2002kv,Calabrese:2002xy,Anderson,Kreiss:2013gw,Sarbach:2012pr}).  We will not dwell on this issue here, but leave it for future work.

In~\cite{Andrade:2015gja} we will show that Minkowski space with a flat Dirichlet wall is linearly stable to purely gravitational perturbations (though we will also find interesting instabilities in other contexts).  A different sort of linearized analysis suggests that flat walls remain stable when exposed to simple non-perturbative processes involving neutral black holes.   In analogy with electrostatics, positive mass sources near a Dirichlet wall must induce negative-mass image-sources behind the wall (see Fig.~\ref{fig:Image}).  Since the gravitational field of a negative mass is repulsive, any black hole is repelled from its image and thus from the wall itself.  As a result, black holes in boxes should oscillate stably about the center of the box.  This expectation will be verified explicitly in \cite{Andrade:2015gja}.

However, giving the the black hole an electric charge causes the image to acquire a charge of the opposite sign.  The gravitational repulsion is then partly cancelled by the electrostatic attraction.  In the extreme limit these forces cancel exactly cancel and further analysis is needed.

We investigate this setting below, exploiting the fact that the cancelation of forces described above carries over to the full non-linear theory in the form of a family of static extremal black holes.
As shown in section \ref{sec:adiabatic}, this is precisely analogous to the cancellation of forces between extremal black holes.  Indeed, our construction is the natural extension of that of 
Majumdar and Papapetrou~\cite{Majumdar:1947eu,Papapetrou} to solutions with Dirichlet walls.
The resulting static black holes provide a family of equal-energy solutions known as a moduli space.  
We then compute the moduli space metric i.e., the kinetic energy associated with motion through this space of solutions.
Techniques for such computations were developed by Ferrel and Eardley to study colliding Reissner-Nordstr\"{o}m back holes~\cite{Ferrell:1987gf} and subsequently refined by Michelson and Strominger to study supersymmetric black holes in five-dimensional ${\cal N} = 1$ supergravity~\cite{Michelson:1999dx}.  The main subtlety in each method is the treatment near the horizon, where a naive calculation gives rise to divergences.  Ferrell and Eardley motivate their regulator by considering lumps of extremal dust and noting that a singular limit recovers the Reissner-Nordstr\"{o}m metric in Majumdar--Papapetrou coordinates (see~\eqref{eq:staticmetric} below).  In contrast, Michelson and Strominger forgo this regulator and instead work directly with delta functions sources.  Through clever manipulations they arrive at a finite expression for the moduli space metric without an explicit regulator.

Applying analogous methods in our setting (see appendix \ref{app:BulkIntegral}) yields a surprising result.  Using
$h$ to parametrize the separation between the black hole and the Dirichlet wall, one finds the moduli space metric
\begin{equation}
\label{SimpleMSM}
ds^2 = \frac{m}{2} \left[ dx^2 + dy^2 + \left(1 - \frac{3m}{h}
- \frac{m^2}{h^2} - \frac{m^3}{2h^3} \right)dh^2 \right].
\end{equation}
One sees that the coefficient of $dh^2$ becomes negative close to the wall, so that the black hole has negative kinetic energy.  One might say that the center-of-mass motion of the black hole contains a ghost for small $h$.  The computation is straightforward (but tedious) and is displayed in appendix \ref{app:BulkIntegral} for those wishing to check the details.

The result \eqref{SimpleMSM} is surprising enough that one should ask whether the techniques of \cite{Ferrell:1987gf} and \cite{Michelson:1999dx} necessarily give the correct result in the present context.  Indeed, though supersymmetry provides a strong constraint in the actual case studied in \cite{Michelson:1999dx}, we have found no proof in the literature fully justifying the treatment of divergences associated with the horizon.

We therefore dedicate the majority of our work below to a complementary approach using Hamiltonian methods.  Because they represent the total energy of a gravitational system as a boundary term at infinity, far from any possible divergence at the horizon, Hamiltonian methods are explicitly finite and gauge invariant.  However, as we will see, they highlight the need for a boundary condition at the horizon to make the problem well-defined.  Below, we make the technically simple choice to cut off the near horizon region of the black hole throat by introducing second (topologically spherical) Dirichlet wall.

Since the throat of an extremal black hole is infinitely deep, the original reasoning of Ferrell and Eardley suggests that moving the cutoff surface arbitrarily close to the horizon should not depend on the details of our boundary condition.  But we will in fact find a moduli space metric that differs from \eqref{SimpleMSM}.  We therefore review this intuition more critically in section \ref{sec:adiabatic}.  In the end we will suggest -- even in the absence of Dirichlet walls -- that the moduli space approximation (also known as the adiabatic approximation) is ill-defined for classical extreme black holes due to the presence of long-lived excitations near the horizon; these are the excitations that were associated with turbulence in \cite{Yang:2014tla}.

Interestingly,  the additional term provided by our Dirichlet wall regulator fails to remove the negative eigenvalue for black holes sufficiently close to the wall. Indeed, as we discuss below, the additional term contains a pole, and is associated with further pathologies of its own.   Obtaining negative kinetic energies from both techniques suggests that this is the qualitatively correct physics and that kinetic energies can indeed become negative near the wall.  This in turn indicates an instability to the black hole acquiring a large velocity.  Unless some form of ghost condensation occurs, the Hamiltonian will be unbounded below.

Such behavior would of course be forbidden in the absence of Dirichlet walls.  For example, in flat space the positivity of kinetic energy on moduli space follows from the positive energy theorem of Gibbons and Hull~\cite{Gibbons:1982fy}.  By adapting the technology developed by Witten in~\cite{Witten:1981mf}, Gibbons and Hull show that $E\ge |Q|$ where $E$ is the ADM energy and $Q$ is the total electric charge.  Since $Q$ is also the energy of black holes at rest, the kinetic energy $E - Q$ is non-negative.

To investigate whether our Hamiltonian is bounded below, and also whether our negative kinetic energies are in tension with the arguments of \cite{Gibbons:1982fy}, we attempted to adapt the proof of~\cite{Witten:1981mf,Gibbons:1982fy} to a spacetime with a Dirichlet wall.  These attempts, however, were unsuccessful, leaving open the possibility that the theory is truly unstable.  The difficulties with the proof may be connected to the fact that Dirichlet boundary conditions necessarily break supersymmetry~\cite{vanNieuwenhuizen:2005kg}\footnote{This fact is easily seen in the AdS context from the well-known result that holographic renormalization performed by imposing a finite-distance cut-off requires a cosmological-constant counter-term in the dual CFT (which would be forbidden if supersymmetry were preserved). We thank Will Donnelly and Joe Polchinski for discussions on this point.}.  We briefly review the Witten style of proof and discuss the obstruction in Appendix~\ref{app:PET}.  The essential point is that we are unable to construct the Green's function necessary to solve $\slashed\nabla \epsilon = 0$.

The organization of the rest of the paper is as follows.  In section~\ref{sec:adiabatic} we construct a family of static extremal black holes parameterized by their mass and distance from the Dirichlet wall and derive a simple expression for their kinetic energy in the low velocity limit.  Then, in section~\ref{sec:calcQ} we perform the integrals necessary to obtain the kinetic energy in closed from.  Finally, in section~\ref{sec:Discussion} we discuss the physical significance of our results.  We collect several technical results in further appendices.

\section{Extremal black holes in the adiabatic limit} \label{sec:adiabatic}

This section develops the technology we will need to compute the kinetic energy of extreme black holes approaching a flat Dirichlet wall, where we use a second (roughly spherical) Dirichlet wall to regulate the calculation near the horizon and to impose a clean boundary condition.  We first construct a moduli space of static solutions and then find an approximate solution for a slowly moving black hole.  Hamiltonian methods allow the kinetic energy of this solution to be written in an integral form that will be explicitly evaluated later in section \ref{sec:calcQ}.

\subsection{The Moduli Space of Static Solutions} \label{sec:ModuliSpace}

Recall that the Einstein--Maxwell equations admit the Majumdar--Papapetrou (MP) family of solutions~\cite{Majumdar:1947eu,Papapetrou} which take the form
\begin{align} \label{eq:staticmetric}
\bar{g}_{ab} dx^a dx^b = - \bar\psi^{-2} dt^2 + \bar\psi^{2} (dz^2 + d\rho^2 + \rho^2 d\phi^2)
 \qquad  \bar{A}_a dx^a = \bar\psi^{-1} dt \, ,
\end{align}
where $\bar \psi$ is any function satisfying
\begin{align}
\label{eq:Laplace}
\grad^{\, 2} \bar\psi = 0 \, ,
\end{align}
and $\grad$ is the flat space (spatial) covariant derivative.  The overbar denotes an exact solution at zero velocity and will be dropped for $v\neq 0$.

We now impose boundary conditions on $\bar \psi$ at $z=0$.  Let $\bar\Sigma$ be a constant $t$-hypersurface in the half-space $z\ge 0$.  The boundary $\partial\bar\Sigma$ consists of two pieces, the surface $z=0$ which we call $\partial\bar\Sigma_0$ and the asymptotic boundary $\partial\bar\Sigma_\infty$.  We require
\begin{align}
\bar\psi=1 \ {\rm at} \ \partial\bar\Sigma_0, \qquad {\rm and} \ \bar\psi = 1 + O(r^{-1}) \ {\rm near} \ \partial\bar\Sigma_\infty,
\end{align}
where $r=\sqrt{\rho^2+z^2}$.  These boundary conditions ensure that the solution satisfies
\begin{itemize}
\item
a Dirichlet boundary condition $h_{AB} = \eta_{AB}$, where $h_{AB}$ is the induced metric on the surface $z=0$ (we use $A,B$ for indices along the wall) and $\eta_{AB}$ is the 2+1 Minkowski metric,
\item
a conducting boundary condition ($E_\parallel=0, B_\perp = 0$) on the vector field $A_a$ at $z=0$
\item
an asymptotically flat boundary condition far from the wall.
\end{itemize}
If we allow $\bar \psi$ to be singular at a single point in $\bar\Sigma$, we find the family of extremal black hole solutions
\begin{align} \label{eq:psiWall}
\bar\psi = 1 + \frac{m}{|\vec x- \vec x_+|} - \frac{m}{|\vec x - \vec x_-|},
\end{align}
where $ \vec x_\pm = \pm h \hat z $.  The integration constant $m$ is the mass of the black hole (which is equal to the electric charge) and $h$ parametrizes the separation between the black hole and the wall.  The singular point $(\rho=0,z=h)$ is the location of the black hole horizon.

An interesting property of the solution~\eqref{eq:psiWall} is that for $h/m < 1/2$ it contains trapped surfaces {\it outside} the black hole event horizon.  This may be seen by defining the radial coordinate
\begin{align}
r_+ := | \vec x - \vec x_+| = \sqrt{\rho^2 +(z-h)^2} \, ,
\end{align}
so that the area element $\sqrt{\omega}$ on (topologically spherical) spacelike surfaces of constant $r_+$ takes the form
\begin{align}
\label{rtom}
\sqrt{\omega} =  r_+ \psi = m + \left( 1- \frac{m}{2h}\right) r_+ + O(r_+^2) \, .
\end{align}
So for $h/m<1/2$ the area of such spheres near the horizon decreases as we move outward.  For a static solution this implies the existence of outer-trapped surfaces outside the black hole -- a feature that cannot arise in asymptotically flat spacetimes satisfying the null energy condition without Dirichlet walls (see e.g. proposition 12.2.2 in~\cite{wald1984general}).  This suggests that our Dirichlet wall may also lead to other features associated with violations of the null energy condition.  We will confirm this suspicion below when we find negative kinetic energies.    Interestingly, while we have found no trapped surfaces outside the black hole for $h> m/2$, the metric \eqref{SimpleMSM} implies we will nevertheless find that the kinetic energy becomes negative at $h/m\approx 3.3$, well away from $h/m=1/2$, and the same will remain true with our Dirichlet regulator at $r=a$.

\subsection{Moving solutions and the great divide}
\label{sec:GD}

The geometric feature just described will turn out to force a division of our moduli space into two disconnected pieces when we add the additional Dirichlet boundary condition at $r=a$. But before discussing this ``great divide'' in detail, let us briefly recall just why such a regulator is needed at all.

The point, of course, is that computing the kinetic energy requires approximate solutions in which the black hole has a small velocity $v$, so that the position of the black hole is time dependent. This entails promoting $\bar \psi(\vec x)\to \psi(t,\vec x)$. But since since $\bar \psi$ diverges as $\vec x \to \vec x_+$, simply replacing $\vec x_\pm \to \vec x_\pm \pm v t$ does not lead to a controlled expansion in $v$.  Some regularization scheme is then needed to render $\psi$ and its derivatives bounded so that we may meaningfully expand in $v$.

\begin{figure}
\centering
\includegraphics[width=0.7 \textwidth]{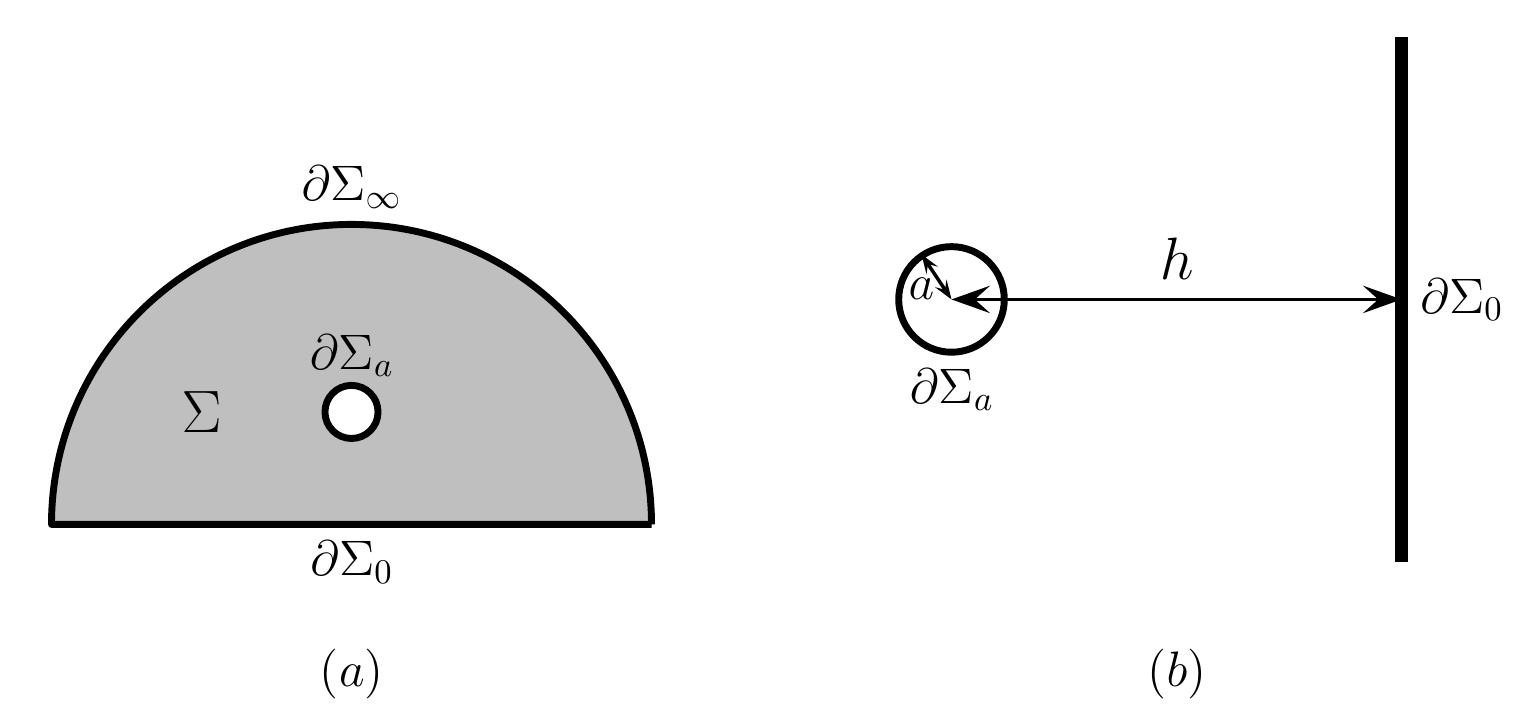}
\caption{(a) A compactified representation of the Cauchy surface $\Sigma$ and the three components of the boundary.  $\partial\Sigma_0$ is the Dirichlet wall, $\partial\Sigma_a$ is the cutoff surface (on which we also impose a Dirichlet boundary condition), and $\partial\Sigma_\infty$ is spacelike infinity at which we require asymptotically flatness.  (b) The black hole cutoff surface is a sphere centered at $(z=h,\rho=0)$ with radius $a$.}
\label{fig:Boundary}
\end{figure}

\begin{figure}
\centering
\includegraphics[width=0.4 \textwidth]{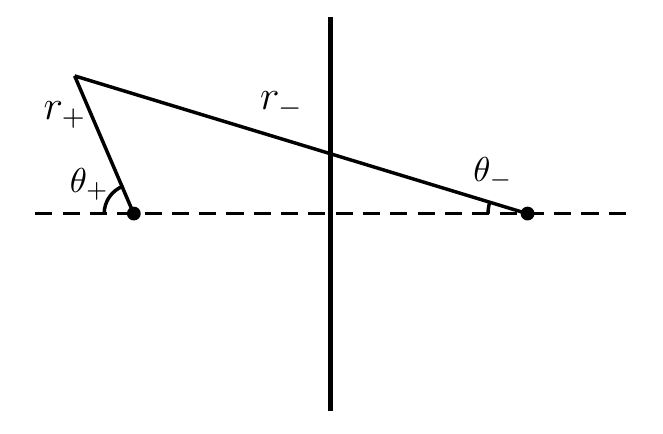}
\caption{A sketch of the polar coordinates $(r_\pm,\theta_\pm)$ used in the text.}
\label{fig:Coord}
\end{figure}

We choose to introduce a radial cutoff $a$, calculate the energy for small $v$, and to then remove the cutoff by taking $a\to 0$.  Let $\Sigma$ be the $t=0$ slice of our non-static spacetime and let $\partial\Sigma$ be the boundary of $\Sigma$.  The boundary $\partial\Sigma$ now has three components (see Fig.~\ref{fig:Boundary}): The asymptotic boundary $\partial\Sigma_\infty$  far from the wall, the wall $\partial\Sigma_{0}$, and the radial cutoff surface $\partial\Sigma_a$.  We take $\partial\Sigma_a$ to be the codimension-two surface $r_+ = a$ and require the induced metric to be a (round) $S^2 \times \mathbb{R}$:
\begin{align} \label{eq:cutoffmetric}
\left. ds^2 \right|_{\partial\Sigma_a} = - dT^2 + R^2 d \Omega^2,
\end{align}
where $d\Omega$ is the line element on the unit sphere and $R$ is some fixed radius.
The full set of boundary conditions on $\psi$ now reads
\begin{align} \label{eq:psicutoff}
\bar\psi=1 \ {\rm at} \ \partial\bar\Sigma_0, \qquad \ \bar\psi = 1 + O(r^{-1}) \ {\rm near} \ \partial\bar\Sigma_\infty,
 \qquad  {\rm and} \ \psi =  R/a \ {\rm at} \ \partial\Sigma_a.
\end{align}
Following \cite{morrison1989potential}, the solutions satisfying~\eqref{eq:psicutoff} can be approximated by the sequence of harmonic functions $\{ \psi_k \}$ where (see Fig.~\ref{fig:Coord})
\begin{align} \label{eq:psiExpansion}
\psi_k =1 + \left(\frac{R}{a}-1\right) \sum_{\ell=0}^k  a^{\ell+1} B^{(k)}_\ell \left( \frac{P_\ell(\cos(\theta_+))}{r_+^{\ell+1}} - (-1)^\ell \frac{P_\ell(\cos(\theta_-))}{r_-^{\ell+1}} \right).
\end{align}
The multipole moments $B^{(k)}_\ell$ in the above expression are given by
\begin{align}
B^{(k)}_\ell = \delta_{0,\ell} + (-\lambda)^\ell \sum_{p=1}^{k+1} \lambda^p \frac{(A_{p-1})^\ell}{(A_p)^{\ell+1}}.
\end{align}
Here $\lambda= a/(2h)$ and the $A_p$ are given by
\begin{align}
A_p = \frac{\left(1+\sqrt{1-4\lambda^2}\right)^{1+p} - \left(1-\sqrt{1-4\lambda^2}\right)^{1+p}}{2^{1+p} \sqrt{1-4\lambda^2} }.
\end{align}

We can now bound the error function $\Delta\psi_k := \psi-\psi_k$.  First note that $\Delta\psi_k$ vanishes on $\partial\Sigma_0,\partial\Sigma_\infty$.  By the maximum principle of the Laplace equation, $|\Delta\psi_k|$ must achieve its maximum value on $\partial\Sigma_a$.  In the limit of small $a$ we have $B^{(k)}_\ell\sim \lambda^{\ell+1}$ and it is straightforward to derive the bound
\begin{align} \label{eq:psiBound}
\max\left(|\Delta\psi_k| \right) \le \frac{R }{h} \lambda^{k+1} + O(\lambda^{k+2}) < \alpha \lambda^{(1-\delta)(k+1)} 
\end{align}
for any $k\ge 1$.  Here $\delta$ may be arbitrarily small so long as $\alpha$ is appropriately large.  Since $\lambda\le 1/2$, the sequence $\{\psi_k\}$ converges uniformly to $\psi$.

In section~\ref{sec:calcQ} we will also need to approximate derivatives of $\psi$.  Arguments given in Appendix~\ref{app:convergence} show that for $k\ge 1$
\begin{align}\label{eq:psiBoundn}
\max\left( | \grad_{i_1}\dots \grad_{i_n} \Delta \psi_k | \right) <
\frac{\beta \lambda^{(1-\delta)(k+1)}}{(h/6)^{n}}.
\end{align}

Since multipole corrections carry no mass or charge, we can write the exact mass (and charge) $m$ of the static solution $\psi$ as
\begin{align}\label{eq:mdef}
m= \left(R-a\right) \lim_{k\to \infty}  B_{0}^{(k)}.
\end{align}
Inverting this expression and expanding for small $a$ gives
\begin{align} \label{eq:RofMandA}
R = m + a \left(1  - \frac{m}{2h}\right) + O(a^4/h^4).
\end{align}

We may use \eqref{eq:RofMandA} to eliminate $R$ in a favor of $m$. and write $\psi_1$ explicitly in the form
\begin{align}\label{eq:psiapprox}
\psi_1 &= 1+(R-a)\left[\left( 1 + \lambda + \frac{\lambda^2}{1-\lambda^2}\right) \left(\frac{1}{r_+} - \frac{1}{r_-}\right) - a  \left( \lambda^2+ \frac{\lambda^3}{(1-\lambda^2)^2}\right) \left(\frac{\cos(\theta_+)}{r_+^2} - \frac{\cos(\theta_-)}{r_-^2}\right) \right]
\nonumber \\
&= 1+ m \left( \frac{1}{r_+} - \frac{1}{r_-} \right) - \frac{m a^3}{(2h)^2} \left( \frac{\cos(\theta_+)}{r_+^2} + \frac{\cos(\theta_-)}{r_-^2}\right) + \dots,
\end{align}
where  $\vec x_\pm=\pm h \hat z$, $(r_\pm,\theta_\pm)$ are polar coordinates centered at $\vec x_\pm$ as in figure \ref{fig:Coord}, and the $\dots$ in the second line represent terms which are subleading as $a\to 0$.

However, \eqref{eq:RofMandA} turns out to have a more fundamental implication.
For $h > m/2$ with $a$ small and positive, it is clear that $R>m$ as one expects.  But for $h< m/2$ one finds $R < m$.  This is another manifestation of the geometric result discussed around \eqref{rtom}.  But since $R$ is fixed by our choice of induced metric at $r=a$  -- i.e., as a boundary condition -- this requires us to use different Dirichlet regulators on each side of $h=m/2$.  We are unable to regulate the entire moduli space at once.  This is the ``great divide'' mentioned earlier.  It suggests that the two pieces may fail to join smoothly at $h=m/2$ even when the regulator is removed.

It is natural to ask if one could do better with another boundary condition.  Rather than investigate specific alternatives, we conclude this section by estimating the dependence of our final results on the particular choice of boundary condition at $r=a$.  One might expect that the infinitely deep throat of the extremal black hole would remove sensitivity to the boundary condition as $a \rightarrow 0$, but this intuition can be tested as follows.  Let $\Delta t$ be the minimum time, as measured at $\partial \Sigma_0$, for a light signal to make a round trip from $\partial \Sigma_0$ to $\partial \Sigma_a$ and back.  This is the timescale on which we expect the boundary condition at  $\partial \Sigma_a$ to become relevant to the motion of the black hole.  We can then calculate $\Delta h$, the change in $h$ over a time $\Delta t$.  The result is
\begin{align} \label{eq:Deltah}
\Delta h = v \Delta t = v \int_h^a \frac{dr}{\dot r} = v \int_h^a dr \bar\psi^2 = mv \left(\frac{m }{a} + 2\log(h/a) \right)+ \dots,
\end{align}
where $\dots$ denote terms that do not diverge as $a\to 0$.

In order to obtain a controlled expansion in which the metric backreaction is small, it will be important below to take $v\to0$ such that $\dot \psi \sim v \grad \psi \sim m v/a^2 $ does not become large; say that $m v/a^2 $ remains less than some constant $c$.  Then \eqref{eq:Deltah} gives
\begin{align}
\label{eq:dh}
\frac{\Delta h}{m a} \lesssim c.
\end{align}
This $\Delta h$ is the distance over which causality protects finite motions on moduli space from dependence on the details of the boundary condition at $r=a$.  And \eqref{eq:dh} shows that $\Delta h \to 0$ as $a\to 0$.  So in this limit causality turns out to provide no protection at all.

This suggests that the adiabatic approximation is not in fact well-defined for classical extreme black holes.  Instead, it will depend on a choice of regulator, perhaps implemented as a boundary condition at the horizon.

Nevertheless, we would like to investigate some specific case in more detail in the hopes of extracting useful qualitative features. To this end, we will carry a detailed calculation with the particular Dirichlet regulator at $r=a$ described above.  We will compute
\begin{align} \label{eq:Tdef}
T = \lim_{a \to 0} \left( \lim_{v\to0} \frac{E - m}{v^2} \right) ,
\end{align}
were $E$ is the total energy and $m$ is the energy of the static solution. In addition to an ADM-like boundary term at infinity, we expect $E$ to have contribution from the Brown-York stress tensor \cite{Brown:1992br}  at the wall.  At least with our choice of regulator, $v^2 T$ is the leading order approximation to the kinetic energy of the black hole.

\subsection{The adiabatic expansion} \label{sec:Adiabatic}

In order to study the kinetic energy $v^2T$, we construct approximate solutions for small $v$.   We wish to work in the adiabatic limit, in which we can associate our solution with a trajectory through the moduli space of (regulated) static solutions.  In practice this means that we replace the boundary condition~\eqref{eq:psicutoff} with
\begin{align} \label{eq:psicutoffoft}
\bar\psi=1 \ {\rm at} \ \partial\bar\Sigma_0, \qquad \ \bar\psi = 1 + O(r^{-1}) \ {\rm near} \ \partial\bar\Sigma_\infty,
 \qquad  {\rm and} \ \psi =  R/a(t) \ {\rm at} \ \partial \Sigma_a(t).
\end{align}
We take the boundary $\partial\Sigma_{a(t)}$ to be the surface $r_+(t) = \sqrt{\rho^2 +(z-h(t))}$ with $a(t)$ is defined in terms of $h(t)$ by using~\eqref{eq:mdef} and requiring that $R,m$ be time independent.  At time $t=0$ we let $\dot h = -v$, where an over-dot denotes a $t$-derivative.  We find
\begin{align}
\dot a = \frac{ m a v}{h(2 h- m)} + O(a^4) \, .
\end{align}
At each time $t$ we take $\psi$ to solve the Laplace equation \eqref{eq:Laplace} with boundary conditions~\eqref{eq:psicutoffoft}.  This defines $\psi(x, t)$.

We will need the time derivative $\dot \psi$ to compute the kinetic energy below.  Since $\dot \psi$ is a solution to the Laplace equation, it is uniquely determined by its boundary data.  Using~\eqref{eq:psicutoffoft}, we can write down boundary data for $\dot \psi$:
\begin{align} \label{eq:psidotcutoffoft}
\dot\psi = 0 \ {\rm at} \ \partial\bar\Sigma_0, \qquad \dot\psi = O(r^{-1}) {\rm near} \ \partial\bar\Sigma_\infty, \qquad \dot \psi =\dot a \partial_a\left(  \frac{R}{a} \right) - \left. \dot r_+  \partial_{r_+} \psi \right|_{r_+=a} \, {\rm at} \ \partial \Sigma_{a(t)}.
\end{align}

The first term at  $\partial\Sigma_{a(t)}$ is the time derivative of the boundary condition on $\psi$, while the second term accounts for the fact that $\partial\Sigma_{a}$ is moving through space.  The partial derivative $\partial_{r_+}$ is taken after expressing $\psi$ in terms of the coordinates $(r_+,\theta_+)$ from figure \ref{fig:Coord} using the relations
\begin{align}
r_- = \sqrt{(r_+ \sin(\theta_+))^2 + (r_+ \cos(\theta_+) + 2h)^2}, \qquad \tan(\theta_-) = \frac{r_+ \sin(\theta_+)}{r_+ \cos(\theta_+) + 2h}\, .
\end{align}
We may then find $\dot r_+$ by solving
\begin{align}
\frac{d}{dt} \left( r_+(t)^2 \sin^2(\theta_+)+ [r_+(t) \cos(\theta_+) + v t]^2 - a(t)^2 \right) = 0,
\end{align}
and setting $r_+ = a$.  This yields
\begin{align}
\left. \dot r_+ \right|_{r_+=a} = \dot a - v \cos(\theta_+).
\end{align}
Below, we will approximate $\dot \psi$ by a function that is precisely harmonic but only approximately satisfies the boundary conditions~\eqref{eq:psidotcutoffoft}.  We will construct such a function at the end of section \ref{sec:Kinetic}.

For $v \neq 0$ the full spacetime metric and Maxwell potential contain additional components not found in \eqref{eq:staticmetric}.  A general ansatz is
\begin{align} \label{eq:ansatz}
ds^2 &= -  (\psi + \delta \psi + \delta\psi_t)^{-2}  dt^2 + 2 \psi^{-2} \vec R \cdot d\vec x dt+ (\psi + \delta \psi)^{2} (\delta_{ij} + \sigma_{ij}) dx^i dx^j, \nonumber \\
A_a dx^a &= -(\psi + \delta \psi + \delta \psi_A)^{-1} dt + (\vec P - \psi^{-1} \vec R) \cdot d\vec x ,
\end{align}
where $\sigma_{ij}$ is traceless.  We also require the solution to be invariant under the combined transformation $t\to -t$ and $v\to -v$,  and to reduce to~\eqref{eq:staticmetric} when $v=0$.  This means that
\begin{align}
\vec R, \vec P \sim v, \qquad \delta\psi, \delta\psi_t,\delta\psi_A,\sigma_{ij} \sim v^2.
\end{align}
The form of the ansatz was chosen so that $\vec R$ transforms simply under the coordinate transformation $t\to t+ \mu (\vec x)$, namely as $\vec R\to \vec R + \grad \mu$.

By calculating the induced metric and gauge field on the codimension-one surface $r_+(t)=a(t)$ we see that our Dirichlet boundary conditions require
\begin{align} \label{eq:y1BC}
\vec R = \frac{3}{2} v \psi^4 (\hat r\cdot \hat z) \hat r, \qquad \vec P = \frac{1}{2} v \psi^4 (\hat r\cdot \hat z) \hat r
\end{align}
at this surface.  The situation is simpler at $z=0$ where we need only require $\vec R,\vec P$ to be normal to the wall.  Asymptotic flatness further requires that the metric and gauge field satisfy
\begin{align} \label{eq:LargeRBC}
g_{ab} -\eta_{ab} \sim  \frac{1}{r}, \qquad A_a -(dt)_a \sim \frac{1}{r}.
\end{align}
Boundary conditions on the second order quantities $\delta\psi,\delta\psi_t,\delta\psi_A,\sigma_{ij}$ at $\partial\Sigma_0$ and $\partial\Sigma_a$ can be also worked out from our Dirichlet boundary conditions, though they will not be needed below.

\subsection{A simple expression for the kinetic energy}
\label{sec:Kinetic}

We may now compute the kinetic energy $v^2 T$.  As noted in \ref{sec:GD}, in addition to the familiar ADM-like boundary term at infinity, the energy term in $T$ should receive a contribution from the Brown-York stress tensor at the wall.  One could proceed by computing both contributions and manipulating the resulting expressions.  However, we will instead use the Landau-Lifshitz technique~\cite{landau1975classical} to express the total energy as a bulk integral of the graviton stress tensor (see also the closely related approaches of \cite{Abbott:1981ff,Deser:2002jk}).

The basic idea of the Landau-Lifshtiz technique is to note that, about any background solution, the linearized Einstein equations define a tensor $H^{(1)}_{ab}$ that is an identically-conserved tensor with respect to the background covariant derivative $\bar{\nabla}_a$; i.e. $\bar{\nabla}_a H^{(1)ab} =0$ for any linearized fields (whether or not they satisfy the linearized equation of motion).  So if the background has a timelike Killing field $\xi^a$, integrating $\xi^a n^b H^{(1)}_{ab}$ over a Cauchy surface with normal $n^b$ defines a conserved quantity even when the fields are not small.  In other words, we may replace what were formally tangent vectors to the space of solutions with the finite differences between the actual fields in any solution and fields in the background.  One then notes that $\xi^a n^b H^{(1)}_{ab}$ is a total derivative, so as expected this energy can be written as a boundary term\footnote{At least to the order used here, the equivalence of this approach with other standard definitions follows from e.g. \cite{Iyer:1994ys} which, as shown explicitly in \cite{Faulkner:2013ica}, implies that when perturbing around a stationary background, the lowest-order change in the Hamiltonian is the sum of the above Landau-Lifshitz term and the change an appropriate horizon area.  The latter term vanishes in our context since the region between infinity and our wall at $r=0$. is horizon-free.  It then follows that for us the lowest-order change occurs at what we have called second order, and that this is properly computed by Landau-Lifhshitz techniques.}. A major advantage of this approach is that it does not require us to solve the second order equations of motion. In fact, as in~\cite{Ferrell:1987gf,Michelson:1999dx}, we will see that it is not even necessary to solve the linearized equations of motion.

The same idea can be applied to the electric charge.  So we may compute the kinetic energy \eqref{eq:Tdef} using the conserved current
\begin{align} \label{Jdef}
\J_a := \frac{1}{\kappa} \left[ \xi^b H_{ab}^{(1)} - \frac{2}{\psi}\nabla^b F_{ab}^{(1)} \right].
\end{align}
In addition, $\xi$ is the background timelike Killing vector and
\begin{align}
H_{ab} &:= G_{ab} - 8\pi T^{EM}_{ab}  ,
\end{align}
where $\kappa = 8\pi$, and $\xi = -\partial_t $ is the background ($v=0$) timelike Killing vector. The superscript (1) indicates that we keep only terms linear in the metric and Maxwell fields (though these terms can be non-linear in the velocity $v$).  A straightforward calculation gives
\begin{align} \label{J2ndOrder}
&\qquad \qquad \kappa \vec \J = -\partial_t \vec Z , \quad \kappa\J_0 = -\psi^{-4}  \grad  \cdot \vec Z, \cr
&Z_i = 2\psi \partial_i (\psi^{-2} \delta\psi_A) + \frac{1}{2} \partial_j \sigma_{ij} + 2 \delta\psi_t \psi^{-2}  \partial_i \psi - 2 \psi \dot A_i
\end{align}
where we have used the background equations of motion $\grad^2 \psi =0$ and the linearized equations of motion.  One may then verify directly that
\begin{align}
\bar\nabla^a \J_a = 0, \label{Jconservation}
\end{align}
where $\bar\nabla$ is the background covariant derivative associated with setting $v=0$.

The associated charge is
\begin{align} \label{eq:Qdef}
{\cal Q} := - \int_\Sigma \sqrt{\bar \gamma} \bar u^a \J_a ,
\end{align}
where $\bar u^a$ and $\bar \gamma$ are the background unit normal and induced metric on the $\Sigma$ (which was defined above to be the surface $t=0$).  Since \eqref{eq:Tdef} may be written
\begin{align}
T = \lim_{a\to 0} \left( \lim_{v\to 0} \frac{{\cal Q}}{v^2}\right),
\end{align}
we need only compute ${\cal Q}$ to second order in $v$.  It is straightforward to check explicitly that ${\cal Q}$ is conserved to this order using
\begin{align}
\dot {\cal Q} &= \partial_t \left(\frac{1}{8\pi}  \int_{\partial \Sigma} d^2 y \, n \cdot \vec Z \right)  = -  \int_{\partial \Sigma} d^2 y \, n \cdot \vec \J ,
\end{align}
where $n$ is the unit normal to $\partial \Sigma$.  By time reversal symmetry $\vec \J$ can only contain odd powers of $v$, but the terms of order $v$ vanish by the linearized equations of motion.

We now derive a convenient expression for computing ${\cal Q}$.  We begin with the equality
\begin{align} \label{eq:EOM}
H_{00} + \frac{2}{\psi}\nabla^b F_{0b}  = 0 \, ,
\end{align}
which follows from the Einstein--Maxwell equations.
Taking~\eqref{eq:EOM} and putting all of the linear terms on the left hand side gives
\begin{align} \label{eq:EOM2}
\psi^{-4} \grad \cdot \vec Z &= -\frac{3 \dot\psi^2}{\psi^2} + \frac{ | \grad\times \vec P|^2}{\psi^6} -  \frac{4 (\grad\times \vec P)\cdot(\grad\times \vec R)}{\psi^7} +  \frac{9 | \grad\times \vec R|^2}{4 \psi^8}
\cr & \qquad
+ \frac{\partial_i  \left[ 2 \psi^{-1} \dot\psi R_i   +   \partial_{[i} (\psi^{-4} R_{j]} R_{j})   \right]}{ \psi^4} .
\end{align}

To simplify this expression, we must gain greater control over $\vec P, \vec R.$ Using the ansatz~\eqref{eq:ansatz} and expanding the equations of motion to linear order yields~\cite{Ferrell:1987gf}
\begin{align} \label{eq:LinearConstraintsPhi12}
\frac{\grad \times \vec{P}  }{\psi^2} - \frac{\grad \times    \vec{R} }{\psi^3} &= \grad \times \vec K + \grad \Phi_1 \cr
\frac{\grad \times \vec{P}  }{\psi^3} - \frac{3}{4}\frac{ \grad \times \vec{R} }{\psi^4} &=  \grad \Phi_2 ,
\end{align}
where $\Phi_1,\Phi_2$ are integration functions and $\vec K$ is defined by\begin{align}
\grad \cdot \vec K = -\dot\psi \qquad \grad^2 \vec K = 0.
\end{align}
These equations are equivalent to the gravitational momentum constraint $D_i \pi^{ij}=8\pi n_a T^{aj}_{EM}$ and Maxwell equation $\nabla_a F^{a j}=0$, which are the only non-trivial equations of motion at linear order in $v$.

Noting that each of $\vec K, \vec R, \vec P$ are invariant under $\phi \rightarrow \phi + const$ and $\phi \rightarrow - \phi$, their curls $\vec \partial \times \vec K, \vec \partial \times \vec R, \vec \partial \times  \vec P$ must be invariant under $\phi \rightarrow \phi + const$ but {\it odd} under $\phi \rightarrow - \phi$.  As a result, only the $\phi$-components of these curls can be non-zero.  But the symmetry also requires $\Phi_1, \Phi_2$ to be independent of $\phi$, so the $\phi$-components of the gradients  $\vec \partial \Phi_1, \vec \partial  \Phi_2$ must vanish.

It follows that the curl terms and gradient terms in \eqref{eq:LinearConstraintsPhi12} vanish separately. As a result, we must have
\begin{align}\label{eq:LinearConstraints}
\grad \times \vec R &= - 4\psi^3 \grad \times \vec K \cr
\grad \times \vec P &= - 3\psi^2 \grad \times \vec K .
\end{align}
Combining~\eqref{eq:EOM2} and~\eqref{eq:LinearConstraints} (along with the definition $\dot \psi = - \grad \cdot \vec K$) we obtain
\begin{align}
\grad \cdot \vec Z &= -3 \psi^2 \left[ \left(\grad\cdot \vec K \right)^2 + \left| \grad\times \vec K \right|^2 \right]  + \partial_i  \left[  \frac{2 \dot \psi R_i}{\psi}  +   \partial_{[i} (\psi^{-4} R_{j]} R_{j})   \right].
\end{align}
Inserting this expression into~\eqref{eq:Qdef} gives
\begin{align}
{\cal Q} &= -\frac{3}{\kappa} \int_\Sigma d^3 \vec x \,  \psi^2 \left[ \left(\grad\cdot \vec K \right)^2 + \left| \grad\times \vec K \right|^2 \right]  + \frac{1}{\kappa} \int_{\partial \Sigma} \hat n_i  \left(  \frac{2 \dot \psi R_i}{\psi}  +   \partial_{[i} (\psi^{-4} R_{j]} R_{j})   \right) ,
\end{align}
where $\hat n$ is the outward pointing flat space, unit normal to $\partial\Sigma$.  The second term in the surface integral vanishes by the Dirichlet boundary condition which requires that $\vec R$ is normal to the $\partial \Sigma$.  The first term vanishes everywhere except on the cutoff surface (for which $\dot \psi\ne0$).  We may thus write
\begin{align}\label{eq:Qsimple}
{\cal Q} &= -\frac{3}{8\pi} \int_\Sigma d^3 \vec x \,  \psi^2 \left[ \left(\grad\cdot \vec K \right)^2 + \left| \grad\times \vec K \right|^2 \right]  + \frac{3 v R^3}{8\pi a} \int_{\partial\Sigma_a}  d\Omega  \,  \dot\psi   \cos(\theta_+) ,
\end{align}
without having to solve the linearized constraints.

Computing $\cal Q$ analytically requires us to write down closed form approximations of $\grad \cdot \vec K$ and $|\grad \times \vec K|$ as we did for $\psi$ above.  Such approximations are obtained by replacing $\vec K$ with
\begin{align} \label{eq:Kapprox}
\vec K_1 = \left[-m  \left(v+\frac{v a^3}{2h^3} - \frac{3 a^2 \dot a}{4h^2}\right) \left( \frac{1}{r_+} + \frac{1}{r_-} \right) + \frac{m v a^3 }{4h^2} \left( \frac{\cos(\theta_+)}{r_+^2} - \frac{\cos(\theta_-)}{r_-^2}\right)  \right] \hat z .
\end{align}
An explicit calculation reveals that $\dot \psi_1 := -\grad \cdot \vec K_1$ satisfies the boundary conditions~\eqref{eq:psicutoffoft} up to terms of order $O(a)$.  In appendix~\ref{app:Kvec}, we also show that $|\grad \times (\vec K - \vec K_1)|\sim O(a)$.  As a result,
we will see in section~\ref{sec:calcQ} below that replacing $(\psi,\vec K)\to (\psi_1,\vec K_1)$ gives us sufficient accuracy to compute the kinetic energy~\eqref{eq:Tdef} in the $a\to 0$ limit.

\section{Computing the kinetic energy} \label{sec:calcQ}

As a warmup we now calculate the kinetic energy without the wall at $z=0$ and obtain the expected answer $mv^2/2$.  In the absence of the wall the functions $\psi,\dot\psi,\vec K$ are precisely
\begin{align}
\psi = 1 + \frac{m}{r}, \qquad  \dot \psi = \frac{m v \cos(\theta) }{r^2}, \qquad \vec K =  \frac{mv}{r} \hat z
\end{align}
where $r = \sqrt{\rho^2 + (z - v t)^2}$ and $m=R-a$ so that $\psi_{\partial\Sigma_a} = R/a$.  Inserting these expressions into~\eqref{eq:Qsimple} gives
\begin{align}\label{eq:Qnowall}
{\cal Q} &= -\frac{3}{2}\int_a^\infty dr \frac{m^2 v^2 (m+r)^2}{r^4} + \frac{3 v (m+a)^3}{4 a} \int d(\cos(\theta))\frac{m v \cos^2(\theta)}{a^2}\cr
&= -\frac{m v^2}{2} \left( \frac{R^3}{a^3} - 1 \right)+ \frac{m v^2}{2} \left(\frac{R^3}{a^3} \right) = \frac{m v^2}{2},
\end{align}
as expected.  Note that both terms in~\eqref{eq:Qsimple} diverge as $a\to 0$, though the divergences cancel and leave a finite answer.  Interestingly, for this case our Dirichlet regulator at $r=a$ gives precisely the same result as the techniques of \cite{Ferrell:1987gf} and \cite{Michelson:1999dx}.

We now return to the case with the wall.  First we consider the boundary term in~\eqref{eq:Qsimple} with $\vec K$ replaced by $\vec K_1$. From \eqref{eq:Kapprox} we find $\dot \psi_1 := -  \grad \cdot \vec K_1$. Performing the angular integral then gives
\begin{align}\label{eq:BoundaryIntFinal}
\frac{3 v R^3}{8\pi a} \int_{r_+=a}  d\Omega  \,  \dot\psi   \cos(\theta_+) =  \frac{m v^2}{2} \left( \frac{R^3}{ a^3} + \frac{m^3 (m+h) }{4 h^3 (h-m/2) } \right) + O(a).
\end{align}
As a simple check on this result, note that if we take $h\to \infty$ we recover the boundary term that appeared in~\eqref{eq:Qnowall}.  As with $\psi_k$ we could define a series of functions $\dot\psi_k$ which include higher multipole corrections. Power counting shows that replacing $\dot \psi_1$ with $\dot \psi_k$ with $k\ge 2$ would not change any of the terms shown above.  So the approximation $\vec K_1$ suffices to compute this term in the limit $a \rightarrow 0$.

Evaluating the bulk term is somewhat more involved.  After again replacing $\vec K$ by $\vec K_1$, we make the decompositions $\psi_1 = \psi_0+ \delta \psi$ and $\vec K_1 = \vec K_0+ \delta \vec K$. Let us first study terms involving $\delta\psi,\delta\vec K$.  All but two of these terms vanish by simple power counting.  The remaining terms are
\begin{align} \label{eq:Qdelta}
&-\frac{3}{8\pi} \int d^3 \vec x \, \psi_0^2 \, 2 \left[ ( \grad\cdot \vec  K_0) ( \grad\cdot \delta\vec K) +  ( \grad\times \vec K_0)\cdot ( \grad\times \delta\vec K ) \right]   = - \frac{m^4(m+4h) v^2}{8h^3(h-m/2)} + O(a) \nonumber \\
&-\frac{3}{8\pi} \int d^3 \vec x \, 2\psi_0 \, \delta\psi \left[ ( \grad\cdot \vec  K_0)^2 +  | \grad\times \vec K_0|^2 \right]   = O(a) ,
\end{align}
where we have now evaluated these terms by using power counting to show that, in the first line, only the leading order term contributes as $a\to 0$ and, in the second line, only the spherically-symmetric part of the term in square brackets can contribute.  The second line then vanishes as $a \to 0$ since $\delta \psi$ contains no monopole term. For the same reason, further improving the approximation by passing to $\vec K_n$ for $n > 1$ gives no effect as $a\to 0$.

It remains to calculate the terms involving only $ \psi_0$ and $\vec K_0$.  Here it is convenient to make contact with the technique developed in~\cite{Ferrell:1987gf,Michelson:1999dx}.  Let us first rewrite $\psi_0,\vec K_0$ as
\begin{align} \label{eq:psi0K0}
\psi_0 = 1 + \sum_{A=\pm} \frac{m_A}{r_A}, \qquad    \vec{K}_0 = \sum_{A=\pm} \frac{m_A}{r_A} \vec{v}_A,
\end{align}
where $m_\pm = \pm m$ and $\vec v_\pm = \mp v \hat z$.  We then introduce the derivative operators $\grad_\pm$ which act on $r_\pm = |\vec x - \vec x_\pm|$ by taking gradients with respect to $\vec x_\pm$.  We will treat $\vec x_\pm$ as independent parameters until the end of the calculation at which time we set $\vec x_\pm = \pm h \hat z$.  Using this new notation we can write
\begin{align}
(\nabla \cdot K_0)^2 + |\nabla \times K_0|^2 &= \sum_{A,B} ( \vec{v}_A \cdot \vec{v}_B \vec{\partial}_A \psi \cdot \vec{\partial}_B \psi - \vec{v}_A \cdot \vec{\partial}_B \psi \vec{v}_B \cdot \vec{\partial}_A \psi  + \vec{v}_A \cdot \vec{\partial}_A \psi \vec{v}_B \cdot \vec{\partial}_B \psi ) \nonumber \\
        &= v^2( \vec{\partial}_+ \psi \cdot \vec{\partial}_+ \psi + \vec{\partial}_- \psi \cdot \vec{\partial}_- \psi -2 \vec{\partial}_+ \psi \cdot \vec{\partial}_- \psi  ).
\end{align}
After some straightforward algebra we find that this contribution to $\cal Q$ can be written
\begin{align} \label{eq:Q0}
-\frac{3}{8\pi} \int_\Sigma d^3 \vec x \,  \psi_0^2 \left[ \left(\grad\cdot \vec K_0 \right)^2 + \left| \grad\times \vec K_0 \right|^2 \right] &=    - \frac{v^2}{32\pi}  \int_\Sigma \ {\cal D} \psi_0^4  \cr
&=  - \frac{m v^2}{2}  \left(\frac{R}{a}\right)^3 - \frac{v^2}{32\pi}  {\cal D} \int_{z > 0} \psi_0^4  \, ,
\end{align}
where
\begin{align} \label{eq:Dchidef}
{\cal D} = \vec{\partial}_+  \cdot \vec{\partial}_+ + \vec{\partial}_-  \cdot \vec{\partial}_- -2 \vec{\partial}_+  \cdot \vec{\partial}_- \, .
\end{align}
Here we have  exploited the fact that ${\cal D}\psi_0^4$ can be written as a total divergence.  This allowed us to separate the integral on the first line of \eqref{eq:Q0} into a boundary term on $\partial\Sigma_0\cup \partial\Sigma_\infty$ that remains bounded as $a \to 0$ and a boundary term at $\partial\Sigma_a$ that diverges.  The boundary term on $\partial\Sigma_a$ evaluates to the first term on the second line.  The boundary term $\partial\Sigma_0\cup \partial\Sigma_\infty$ can then be written as a bulk integral over the entire region $z>0$ (including the region $r< a$) treating the integrand ${\cal D} \psi_0^4$ as an appropriate distribution at $r=0$. This term is evaluated in Appendix~\ref{app:BulkIntegral}, which also explains the connection of this term to the effective action of~\cite{Ferrell:1987gf} or~\cite{Michelson:1999dx}.\footnote{cf. Eq.~(2.10) in~\cite{Michelson:1999dx} after setting $\lambda=0$.  Note that in~\cite{Michelson:1999dx} the authors work in five spacetime dimensions, whereas we are working in four. See Appendix~\ref{app:BulkIntegral} for additional details.}  The result is
\begin{equation} \label{eq:DchiResult}
- \frac{v^2}{32\pi} \int_{z>0} {\cal D} \psi_0^4 =  \frac{m v^2}{2} \left(1 - U(m/h) \right),  \ \ \ \ {\rm with} \ \ \
U(x) =\frac{1}{2}  \left(6 x + 2 x^2 + x^3\right)  \, .
\end{equation}

Combining~\eqref{eq:BoundaryIntFinal},~\eqref{eq:Qdelta}, and~\eqref{eq:Q0}, the divergent $R^3/a^3$ terms cancel to yield
\begin{align}\label{eq:Qfinal}
{\cal Q} = mv^2\left[ \frac{1}{2} - \frac{1}{4}\left( \frac{6m}{h} + \frac{2m^2}{h^2} + \frac{m^3}{h^3}\right) - \frac{3 m^3}{8h^2(h-m/2)}\right] + O(a).
\end{align}
The final term inside the brackets (the term with the pole at $h=m/2$) is the sole discrepancy between our Dirichlet-regulated calculation and the result one would obtain using the techniques of~\cite{Ferrell:1987gf,Michelson:1999dx}.  Working backward, one may show that this term originates from the dipole correction to $\psi$ appearing in~\eqref{eq:psiapprox}.  This fact strongly suggests that our final result for the kinetic energy does in fact depend on our choice of regulator.

\section{Discussion} \label{sec:Discussion}

Our work above computed the metric on the moduli space for extreme black holes in the presence of a flat Einstein-Maxwell Dirichlet wall using two distinct methods of dealing with divergences at the horizon.  The results indicate two important lessons.  The first stems from the negative eigenvalue of the metric for black holes close to the wall.  Since it appears using either of our methods, the associated negative kinetic energies appear to be a robust feature of the physics, with corresponding implications for stability.

Indeed, let us imagine that we place an extremal black hole in the region with a negative eigenvalue. Then the negative kinetic energy means that conservation of energy allows any process emitting positive-energy gravitational radiation so long as the black hole acquires a non-zero velocity $v$.  Furthermore, the process can continue (with corresponding increases in velocity) at least until $v$ becomes so large that the moduli space approximation breaks down.  Referring to this component of velocity as a ghost, the system must either exhibit some form of ghost condensation (which would not be visible in perturbation theory) or the Hamiltonian will be unbounded below.  We speculate that the latter may in fact be true.  In this direction, we note in appendix~\ref{app:PET} that standard techniques for proving a positive energy theorem~\cite{Witten:1981mf} break down when we include a Dirichlet wall.  The same comments apply to attempts to use techniques from \cite{Gibbons:1982fy} to prove $E \ge Q$.

The second lesson stems from the fact that our two method gave results that differ in detail.  Indeed, adding an explicit cut-off at $r=a$ led to a new term in the moduli space metric beyond the result \eqref{SimpleMSM} that would be obtained using the techniques of \cite{Ferrell:1987gf} or \cite{Michelson:1999dx}.  With a Dirichlet boundary condition at this cut-off surface, we found a particular such term containing a pole at $h=m/2$ that remains after taking $a\to 0$.   We also saw that the origin of this term was the dipole correction which must be made to $\psi$ to satisfy the Dirichlet condition at $r=a$ for finite $a$.  Since dipole corrections should be expected to arise from any other choice of boundary condition at $r=a$, this suggests that other boundary conditions again lead to $a \to 0$ moduli space metrics that differ from \eqref{SimpleMSM}, and that the result may depend in detail on the boundary condition chosen. Our moduli space of extreme black holes would then be ill-defined.

It is natural to ask whether this regulator-dependence arises only in the presence of the Dirichlet wall $\partial \Sigma_0$.  We have therefore also used our Dirichlet regulator to calculate the moduli space metric for the two black hole scattering problem originally treated in~\cite{Ferrell:1987gf}.  For that case we find the Dirichlet-regulated metric to again contain an additional term beyond those appearing in \cite{Ferrell:1987gf}.  The new term is precisely the one appearing in~\eqref{eq:Qfinal} with $m$ replaced by $-m$.   The dependence of the moduli space metric on regulating boundary conditions thus appears to be a general property of moduli spaces for classical extreme black holes.

Indeed, we see no a priori reason to expect a good adiabatic approximation even as $v \to 0$. The point is that the lifetime of the longest-lived quasinormal mode of a classical black hole turns out to diverge as the black hole becomes extreme.  It is precisely this phenomenon that was associated with turbulence in \cite{Yang:2014tla}.    And for any finite $v$ process, it suggests that a sufficiently extreme black hole can be significantly perturbed from any nominal ``ground state.''  It would thus be very interesting to better determine if (and when) the moduli space calculations of \cite{Ferrell:1987gf} correctly approximate the kinetic energy of an interacting pair of slowly-moving extreme black holes.   While numerical studies (see e.g. \cite{ZilhAO:2015xta}) have not yet reached the level where they can be usefully compared with such results, we may hope that they will do so in the near future.

It remains to address the success of \cite{Michelson:1999dx} in comparing its classical moduli space with string-theoretic calculations. If the classical moduli space is ill-defined, why should there be any such agreement?   Recall that the goal of \cite{Michelson:1999dx} work was to compare a classical black hole calculation with the moduli space for a supersymmetric quantum system.  Quantum mechanics should render the black hole spectrum discrete, so that an adiabatic approximation will in fact hold for sufficiently slow processes\footnote{Since the spectrum will generally have exponentially small spacing, this moduli space metric will govern only exponentially slow motions.  The notable exception occurs in supersymmetric contexts where the gap above the highly-degenerate ground state is only polynomially-small.}.  Thus the desired quantum system should indeed admit a well-defined moduli space.   One may thus view the success of \cite{Michelson:1999dx} as suggesting that requiring the moduli space to maintain supersymmetry is sufficient to extract the associated metric from a classical calculation.

Having argued that the $a=0$ moduli space is ill-defined, we pause to note that spacetimes with finite cutoff parameter $a$ are interesting solutions which form a moduli space in their own right (for a given sign of $h-m/2$).  Taking $a$ to be small but finite we may interpret~\eqref{eq:Qfinal} as the kinetic energy of a solution with two Dirichlet surfaces, one flat and one spherical.  Our results clearly show that such boundary conditions admit negative kinetic energy solutions with $E < Q$.

\acknowledgments

It is a pleasure to thank Gary Horowitz, Juan Maldacena, Jorge Santos, Norihiro Tanahashi, Aron Wall, and Helvi Witek for helpful discussions and feedback.  T.A. was supported by the European Research Council under the European Union's Seventh Framework Programme (ERC Grant agreement 307955).  W.K. and D.M. were supported by the National Science Foundation under grant number PHY12-05500 and by funds from the University of California.

\appendix
\section{The Bulk Integral}\label{app:BulkIntegral}

This appendix derives~\eqref{SimpleMSM}, which gives the moduli space metric for extreme Einstein-Maxwell black holes near a Dirichlet wall as computed using techniques of~\cite{Ferrell:1987gf,Michelson:1999dx}.  This method focuses on the effective action
\begin{gather}\label{eq:Seff}
S_{\text{eff}} = \frac{1}{16\pi}  \int d^4 x \sqrt{-g} (R - F_{ab}F^{ab}) + S_{\text{source}} \cr
S_{\text{source}} = \frac{1}{4\pi}  m \left( \int (u_a -    A_a) dx^a_+ \right)
\, .
\end{gather}
The term $S_{\text{source}}$ couples extremal point particles of of mass and charge $m$ with four velocity $u$ to the fields.  In~\cite{Ferrell:1987gf}, the source terms were derived by writing down the action for an extremal dust and taking the dust distribution to be a delta function.  In~\cite{Michelson:1999dx} these terms are reinterpreted as compensating for the fact that the field equations are not satisfied at the timelike singularity behind the black hole horizon.

We are to compute this effective action for the ansatz
\begin{align} \label{eq:ansatz0}
ds^2 &= -  \psi_0^{-2}  dt^2 + 2 \psi^{-2} \vec R \cdot d\vec x dt+ \psi_0^{2} (\delta_{ij} + \sigma_{ij}) dx^i dx^j, \cr
A_a dx^a &= -\psi_0^{-1} dt + (\vec P - \psi_0^{-1} \vec R) \cdot d\vec x \, ,
\end{align}
with $\psi_0$ as given as in~\eqref{eq:psi0K0}.  After a long but straightforward calculation one finds
\begin{align} \label{eq:SeffFinal}
S_{\text{eff}} &= -\frac{1}{32\pi} \sum_{A=\pm,B=\pm} \int dt (\delta^{ij}\delta_{kl} + \delta^i_k \delta^j_l - \delta^i_l \delta^j_k) v_A^k v_B^l \partial_{Ai}\partial_{Bj} \chi \cr
&= - \frac{v^2}{32\pi} {\cal D}\chi\, ,
\end{align}
where we have used the linearized constraints to eliminate $\vec R,\vec P$ from the action,
 \begin{align} \label{eq:Dchidef1}
\chi = \int_{z >0} \psi_0^4 \, ,
\end{align}
and
 \begin{align} \label{eq:Dchidef2}
{\cal D} = \vec{\partial}_+  \cdot \vec{\partial}_+ + \vec{\partial}_-  \cdot \vec{\partial}_- -2 \vec{\partial}_+  \cdot \vec{\partial}_- \, .
\end{align}
This is a four-dimensional analogue of the 5-dimensional result of \cite{Michelson:1999dx}.

The definition \eqref{eq:Dchidef2} agrees with that of \eqref{eq:Dchidef} from the main text. While $\chi$ diverges, ${\cal D} \chi$ may be evaluated by first acting with ${\cal D}$ on the integrand $\psi^4_0$, treating the result as a distribution, and finally performing the integral over all $z > 0$.  With this understanding we see that ${\cal D} \chi$ agrees with the term evaluated in~\eqref{eq:DchiResult}.

To begin the computation of ${\cal D} \chi$, let
\begin{equation}\label{chi W}
    \chi = I_0 + m I_1 + m^2 I_2 + m^3 I_3 + m^4 I_4,
\end{equation}
\noindent where
\begin{equation}\label{IW0}
    I_0 = \int_{z>0} d^3 x
\end{equation}
\begin{equation}\label{IW1}
    I_{1} = 4 \int_{z>0} d^3 x \left[\frac{1}{r_+} - \frac{1}{r_-} \right]
\end{equation}
\begin{equation}\label{IW2}
    I_{2} = 6 \int_{z>0} d^3 x \left[\frac{1}{r^2_+} + \frac{1}{r^2_-} - \frac{2}{r_+ r_-} \right]
\end{equation}
\begin{equation}\label{IW3}
    I_{3} = 4 \int_{z>0} d^3 x \left[\frac{1}{r_+^3} - \frac{1}{r_-^3} + \frac{3}{r_+ r_-^2} - \frac{3}{r^2_+ r_-} \right]
\end{equation}
\begin{equation}\label{IW4}
    I_{4} = \int_{z>0} d^3 x \left[\frac{1}{r_+^4} + \frac{1}{r_-^4} + \frac{6}{r_+^2 r_-^2} - \frac{4}{r_+ r_-^3} - \frac{4}{r_+^3 r_-} \right].
\end{equation}

As advertised above, we evaluate ${\cal D}I_n$ by acting with $\cal D$ on the integrands (\ref{IW1}), (\ref{IW2}), (\ref{IW3}) and (\ref{IW4}) and calculating the integrals afterwards. We take the the black hole and its image to lie on the $z$-axis in positions which we denote by $z_+ > 0$ and $z_- <0$, respectively, and after taking the desired derivatives we set $z_- = - z_+$  as required by our boundary conditions.

In the spirit of~\cite{Ferrell:1987gf,Michelson:1999dx} we integrate over the entire region $z>0$.  as a result, we will often use the identity $\grad^{\, 2}_+ r_+^{-1} = -4 \pi \delta^{(3)}(\vec{x} - \vec{x}_+)$.   We now collect the useful results
\begin{equation}\label{id0 1 2}
     r_+ = r_- , \ \ \ \ \ \partial_z r_+^{-n}  = \frac{n z_+}{r_+^{n+2}} , \ \ \ \ \ \partial_z r_-^{-n}  = -\frac{n z_+}{r_+^{n+2}} , \ \ {\rm at} \ z =  0,
\end{equation}
\begin{equation}\label{id3}
    \int_{z=0} d^2 x r_+^{-n} = \frac{2 \pi}{(n-2)} z_+^{2-n} \ \ \ \ {\rm for}  \ n > 2,
\end{equation}
\begin{equation}\label{id4}
    \grad^{\, 2} r_-^{-2} = 2 r_-^{-4} + 2 r_-^{-1} \delta^{(3)}(\vec{x}-\vec{x}_-),
\end{equation}
\begin{equation}\label{id5}
    \int_{z>0} d^3 x r_+^{-2} r_-^{-4} = \frac{\pi}{2 z_+^3}.
\end{equation}
We emphasize that identities (\ref{id0 1 2}) and (\ref{id5}) hold under the assumptions that $\vec{x}_+ = - \vec{x}_- = z_+ \hat z$, which we impose after computing all the derivatives. We also mention that identities (\ref{id4}), (\ref{id5}) are useful to compute the term ${\cal D} \int r_+^{-2} r_-^{-2}$. Finally, we shall make extensive use of the fact that $\vec{x}_-$, the position of the image, lies outside of the domain of integration of the integrals (\ref{IW1}), (\ref{IW2}), (\ref{IW3}) and (\ref{IW4}), so volume integrals involving $\delta^{(3)}(\vec{x}- \vec{x}_-)$ vanish.

\subsection{Terms with $m^2$}

We calculate now the quantity ${\cal D} I_2$, where
\begin{equation*}
    I_{2} = 6 \int_{z>0} d^3 x \left[\frac{1}{r^2_+} + \frac{1}{r^2_-} - \frac{2}{r_+ r_-} \right]
\end{equation*}

\noindent $\blacksquare$ {\bf term $r^{-2}_+$}
\begin{equation}
    \grad_+^{\, 2} \int_{z>0} r_+^{-2} = \int_{z>0} \grad^{\, 2} r_+^{-2} = -\int_{z=0} \partial_z r_+^{-2} = -2 z_+ \int_{z=0} r_+^{-4} = -\frac{2 \pi}{z_+}
\end{equation}
\begin{equation}\label{}
    \grad_-^{\, 2} \int_{z>0} r_+^{-2} = 0
\end{equation}
\begin{equation}\label{}
    \grad_+ \cdot \grad_- \int_{z>0} r_+^{-2}  = 0
\end{equation}
Hence,
\begin{equation}
    {\cal D} \int_{z>0} r_+^{-2}  = -\frac{2 \pi}{z_+}.
\end{equation}

\noindent $\blacksquare$ {\bf term $r^{-2}_-$}
\begin{equation}
    \grad_-^{2} \int_{z>0} r_-^{-2} = \int_{z>0} \grad^{\, 2} r_-^{-2} = -\int_{z=0} \partial_z r_-^{-2} =  2 z_+ \int_{z=0} r_+^{-4} =  \frac{2 \pi}{z_+}
\end{equation}
\begin{equation}
    \grad_+^{2} \int_{z>0} r_-^{-2} = 0
\end{equation}
\begin{equation}\label{}
    \grad_+ \cdot \grad_- \int_{z>0} r_-^{-2}  = 0
\end{equation}
Thus,
\begin{equation}
    {\cal D} \int_{z>0} r_-^{-2}  =  \frac{2 \pi}{z_+}.
\end{equation}

\noindent $\blacksquare$ {\bf term $r^{-1}_+ r^{-1}_-$}
\begin{equation}
    \grad_+^{\, 2} \int_{z>0} r_+^{-1} r_-^{-1} = - \frac{4\pi}{|\vec x_+ -\vec x_-|} = - \frac{2 \pi}{z_+}
\end{equation}
\noindent where we have used in the last step that $|\vec x_+ -\vec x_-| = z_+ - z_- = 2 z_+$.
\begin{equation}\label{}
    \grad^{\, 2}_- \int_{z>0} r_+^{-1} r_-^{-1} = \int_{z>0} r_+^{-1} \grad^{\, 2} r_-^{-1} = 0
\end{equation}
\noindent since $\vec{x}_-$ lies outside of the domain of integration.
\begin{equation}
    \grad_+ \cdot \grad_- \int_{z>0} r_+^{-1} r_-^{-1} = \int_{z>0} \vec{\nabla} r_+^{-1} \cdot \vec{\nabla} r_-^{-1} =
    -\int_{z=0} r_+^{-1} \partial_z r_-^{-1} =  z_+ \int_{z=0} r_+^{-4} =  \frac{\pi}{z_+}
\end{equation}
So
\begin{equation}\label{}
  {\cal D}  \int_{z>0} r_+^{-1} r_-^{-1} = - \frac{4 \pi}{z_+},
\end{equation}
and\begin{equation}\label{}
    {\cal D} I_2 = \frac{48 \pi}{z_+}.
\end{equation}

\subsection{Terms with $m^3$}

Now we calculate ${\cal D} I_3$, where
\begin{equation*}
    I_{3} = 4 \int_{z>0} d^3 x \left[\frac{1}{r_+^3} - \frac{1}{r_-^3} + \frac{3}{r_+ r_-^2} - \frac{3}{r^2_+ r_-} \right]
\end{equation*}

\noindent $\blacksquare$ {\bf term $r^{-3}_+$}

\begin{equation}
    \grad_+^2 \int_{z>0} r^{-3}_+ = -\int_{z = 0} \partial_z r_+^{-3} = -3 z_+ \int_{z = 0} r_+^{-5} = -\frac{2 \pi}{z_+^2}
\end{equation}
\begin{equation}\label{}
    \grad_-^2 \int_{z>0} r^{-3}_+  = 0
\end{equation}
\begin{equation}\label{}
    \grad_+ \cdot \grad_- \int_{z>0} r^{-3}_+  = 0
\end{equation}

Hence
\begin{equation}\label{}
    {\cal D} \int_{z>0} r^{-3}_+ = -\frac{2 \pi}{z_+^2}
\end{equation}

\noindent $\blacksquare$ {\bf term $r^{-3}_-$}

\begin{equation}
    \grad_-^2 \int_{z>0} r^{-3}_- = -\int_{z = 0} \partial_z r_-^{-3} =  3 z_+ \int_{z = 0} r_+^{-5} =  \frac{2 \pi}{z_+^2}
\end{equation}
\begin{equation}\label{}
    \grad_+^2 \int_{z>0} r^{-3}_-  = 0
\end{equation}
\begin{equation}\label{}
    \grad_+ \cdot \grad_- \int_{z>0} r^{-3}_-  = 0
\end{equation}

Hence
\begin{equation}\label{}
    {\cal D} \int_{z>0} r^{-3}_- = \frac{2 \pi}{z_+^2}
\end{equation}

\noindent $\blacksquare$ {\bf term $r^{-1}_+ r^{-2}_-$}

\begin{equation}
    \grad_+^2 \int_{z>0} r^{-1}_+ r^{-2}_- = - \frac{4 \pi}{|\vec x_+ - \vec x_-|^2} = - \frac{\pi}{z_+^2}
\end{equation}

\begin{eqnarray}
\nonumber
    \grad_-^2 \int_{z>0} r^{-1}_+ r^{-2}_- &=&  \int_{z>0} r^{-1}_+  \grad^{\, 2} r^{-2}_- \\
\nonumber
                                            &=&  -\int_{z=0} r_+^{-1} \partial_z r_-^{-2} - r_-^{-2} \partial_z r_-^{-1}  +
                                            \int_{z>0} r_-^{-2} \grad^{\, 2} r_+^{-1} \\
\nonumber
                                            &=& - \frac{4 \pi}{|\vec x_+ - \vec x_-|^2} + 3z_+ \int_{z=0} r_+^{-5}  \\
                                            &=&  \frac{\pi}{z_+^2}
\end{eqnarray}

\begin{eqnarray}
\nonumber
    \grad_+ \cdot \grad_- \int_{z>0} r^{-1}_+ r^{-2}_- &=&  \int_{z>0} \vec{\nabla} r^{-1}_+ \vec{\nabla} r^{-2}_- \\
\nonumber
                                            &=&  -\int_{z=0}  r_-^{-2} \partial_z r_+^{-1}  - \int_{z>0} r_-^{-2} \grad^{\, 2} r_+^{-1} \\
\nonumber
                                            &=& \frac{4 \pi}{|\vec x_+ - \vec x_-|^2} - z_+ \int_{z=0} r_+^{-5}  \\
                                            &=& \frac{\pi}{3 z_+^2}
\end{eqnarray}

So we have
\begin{equation}
    {\cal D} \int_{z>0} r^{-1}_+ r^{-2}_-  = - \frac{2 \pi}{3z_+^2}.
\end{equation}

\noindent $\blacksquare$ {\bf term $r^{-2}_+ r^{-1}_-$}

\begin{equation}
    \grad_+^2 \int_{z>0} r^{-2}_+ r^{-1}_- =  \int_{z>0} \grad^{\, 2} r^{-2}_+  r^{-1}_- =  -\int_{z=0} r_-^{-1} \partial_z r_+^{-2} - r_+^{-2} \partial_z r_-^{-1}  = -3 z_+ \int_{z=0} r_+^{-5} =  -\frac{2 \pi}{z_+^2}
\end{equation}

\begin{equation}
    \grad_-^2 \int_{z>0} r^{-2}_+ r^{-1}_-  = 0
\end{equation}
\begin{equation}\label{}
    \grad_+ \cdot \grad_- \int_{z>0} r^{-2}_+ r^{-1}_-  = \int_{z>0} \vec{\nabla} r^{-2}_+ \cdot \vec{\nabla} r^{-1}_- =  -\int_{z=0} r_+^{-2} \partial_z r_-^{-1} = z_+ \int_{z=0} r_+^{-5} =  \frac{2 \pi}{3 z_+^2}
\end{equation}

Thus,
\begin{equation}\label{}
    {\cal D} \int_{z>0} r^{-2}_+ r^{-1}_-  = - \frac{10 \pi}{3 z_+^2},
\end{equation}
and we have
\begin{equation}\label{}
    {\cal D} I_3 =  \frac{16 \pi}{z_+^2}.
\end{equation}

\subsection{Terms with $m^4$}

Finally, we compute ${\cal D}I_4$, where
\begin{equation*}
    I_{4} = \int_{z>0} d^3 x \left[\frac{1}{r_+^4} + \frac{1}{r_-^4} + \frac{6}{r_+^2 r_-^2} - \frac{4}{r_+ r_-^3} - \frac{4}{r_+^3 r_-} \right]
\end{equation*}

\noindent $\blacksquare$ {\bf terms $r^{-4}_+$ and $r^{-4}_-$ } \\

The computation of these terms is closely analogous to the one of ${\cal D} r_\pm^{-3}$. We obtain
\begin{equation}\label{}
    {\cal D} \int_{z>0} r_+^{-4} =  - \frac{2 \pi}{z_+^3}
\end{equation}
\begin{equation}\label{}
    {\cal D} \int_{z>0} r_-^{-4} =   \frac{2 \pi}{z_+^3}
\end{equation}

\noindent $\blacksquare$ {\bf term $r^{-1}_+ r_-^{-3}$  }

\begin{equation}\label{}
    \grad_+^2 \int_{z>0} r^{-1}_+ r_-^{-3} = - \frac{4 \pi}{|\vec x_+ - \vec x_-|^3} = - \frac{\pi}{2 z_+^3}
\end{equation}

\begin{eqnarray}
\nonumber
  \grad_-^2 \int_{z>0} r^{-1}_+ r_-^{-3} &=& \int_{z>0} r^{-1}_+ \grad^{\, 2} r_-^{-3} \\
\nonumber
   &=& - \frac{4 \pi}{|\vec x_+ - \vec x_-|^3} - \int_{z=0} ( r_+^{-1} \partial_z r_-^{-3} - r_-^{-3} \partial_z r_+^{-1}  ) \\
\nonumber
   &=&  - \frac{4 \pi}{|\vec x_+ - \vec x_-|^3}  + 4 z_+ \int_{z=0} r_+^{-6} \\
   &=& \frac{3 \pi}{2 z_+^3}
\end{eqnarray}

\begin{eqnarray}
\nonumber
  \grad_+ \cdot  \grad_- \int_{z>0} r^{-1}_+ r_-^{-3} &=& \int_{z>0} \vec{\nabla} r^{-1}_+ \vec{\nabla} r_-^{-3} \\
\nonumber
   &=& \frac{4 \pi}{|\vec x_+ - \vec x_-|^3} - \int_{z=0} r_-^{-3} \partial_z r_+^{-1}  \\
\nonumber
   &=&  \frac{4 \pi}{|\vec x_+ - \vec x_-|^3}  - z_+ \int_{z=0} r_+^{-6} \\
   &=& 0
\end{eqnarray}

Thus,
\begin{equation}\label{}
    {\cal D} \int_{z>0} r^{-1}_+ r_-^{-3} = \frac{\pi}{z_+^3}.
\end{equation}

\noindent $\blacksquare$ {\bf term $r^{-3}_+ r_-^{-1}$  }

\begin{equation}\label{}
    \grad_+^2 \int_{z>0} r^{-3}_+ r_-^{-1} = -\int_{z=0} (r_-^{-1} \partial_z r_+^{-3} - r_+^{-3} \partial_z r_-^{-1}) = -4 z_+ \int_{z=0} r_+^{-6}  =- \frac{2 \pi}{z_+^3}
\end{equation}
\begin{equation}\label{}
    \grad_-^2 \int_{z>0} r^{-3}_+ r_-^{-1} = 0
\end{equation}
\begin{equation}\label{}
    \grad_+ \cdot \grad_- \int_{z>0} r^{-3}_+ r_-^{-1} = - \int_{z=0} r_+^{-3} \partial_z r_-^{-1} =  z_+ \int_{z=0} r_+^{-6}  =  \frac{\pi}{2 z_+^3}.
\end{equation}

Thus,
\begin{equation}\label{}
    {\cal D} \int_{z>0} r^{-3}_+ r_-^{-1} = -\frac{3 \pi}{z_+^3}
\end{equation}

\noindent $\blacksquare$ {\bf term $r^{-2}_+ r_-^{-2}$  }

\begin{eqnarray}
\nonumber
  \grad_+^2 \int_{z>0} r^{-2}_+ r_-^{-2}   &=& - \int_{z=0} (r_-^{-2} \partial_z r_+^{-2} - r_+^{-2} \partial_z r_-^{-2} ) + \int_{z>0} r_+^{-2} \grad^{\, 2} r_-^{-2}\\
\nonumber
   &=& - 4 z_+ \int_{z=0} r_+^{-6} + 2 \int_{z>0} r_+^{-2} r_-^{-4} \\
   &=& - \frac{\pi}{z_+^3}.
\end{eqnarray}
\noindent Here we have used identities (\ref{id4}) and (\ref{id5}).  These identities may also be used to show

\begin{equation}\label{}
    \grad_-^2 \int_{z>0} r^{-2}_+ r_-^{-2} = \int_{z>0} r^{-2}_+ \grad^{\, 2} r_-^{-2} = 2 \int_{z>0} r_+^{-2} r_-^{-4} = \frac{\pi}{z_+^3},
\end{equation}
and
\begin{eqnarray}\label{}
\nonumber
    \grad_+ \cdot \grad_- \int_{z>0} r^{-2}_+ r_-^{-2} &=& -\int_{z=0} r^{-2}_+ \partial_z r_-^{-2} - \int_{z>0} r_+^{-2} \grad^{\, 2} r_-^{-2} \\
    &=&  2 z_+ \int_{z=0} r_+^{-6} - 2 \int_{z>0} r_+^{-2} r_-^{-4} \\
\nonumber
    &=& 0.
\end{eqnarray}

Hence,
\begin{equation}\label{}
    {\cal D}\int_{z>0} r^{-2}_+ r_-^{-2} = 0,
\end{equation}
and
\begin{equation}\label{}
    {\cal D} I_4 = \frac{8 \pi}{z_+^3}.
\end{equation}

\subsection{Final result}

The action of the operator ${\cal D}$ on the various integrals $I_n$ was found to be
\begin{equation}\label{collect}
        {\cal D} I_0 = 0 \ \ \ \ \ {\cal D} I_1 = - 16 \pi \ \ \ \ \ {\cal D} I_2 = \frac{48 \pi}{h} \ \ \ \ \     {\cal D} I_3 = \frac{16 \pi}{h^2} \ \ \ \ \   {\cal D} I_4 = \frac{8 \pi}{h^3} \, ,
\end{equation}
where we have used $z_+ = h$.  Inserting these results into the expression
\begin{align}
- \frac{v^2}{32\pi}{\cal D} \chi = - \frac{v^2}{32 \pi} {\cal D} \sum_n m^n I_n \, ,
\end{align}
gives \eqref{SimpleMSM}.

\section{Difficulties of proving a positive energy theorem}\label{app:PET}

We now review Witten's method of proving a positive energy theorem and show why this technique appears not to extend to settings with Dirichlet walls.

Witten's argument~\cite{Witten:1981mf} consists of two parts.  One first supposes that the spacetime admits a Cauchy surface $\Sigma$ and a spacetime spinor satisfying
\begin{align} \label{eq:Deps}
\slashed \nabla \epsilon= 0, \qquad \epsilon = \epsilon_\infty + O(r^{-1}) \ \ {\rm on} \ \ \Sigma,
\end{align}
where $\epsilon_\infty$ is a constant spinor,  $\nabla_a$ are the spacetime covariant derivatives,  $\gamma^a$ are the Clifford algebra generators satisfying $\gamma^a \gamma^b + \gamma^b \gamma^a = -2 g^{a b}$, and
$\slashed \nabla = \gamma^i \nabla_i$ with an index $i$ denoting a corresponding object projected into the surface $\Sigma$.  A calculation then shows
\begin{align} \label{eq:Ege0}
(\epsilon_\infty \epsilon_\infty^\dagger)  E \ge \int_\Sigma |\nabla \epsilon|^2 \ge 0,
\end{align}
where $E$ is the energy associated with a future-directed asymptotic Killing field proportional to $\bar \epsilon_\infty \gamma^a \epsilon_\infty$.  The inequality~\eqref{eq:Ege0} follows after rearranging
\begin{align}
\int_\Sigma \epsilon^\dag \slashed \nabla^2 \epsilon = 0,
\end{align}
and assuming the dominant energy condition.

It is then argued that a solution to~\eqref{eq:Deps} always exists for asymptotically flat spacetimes.  This is shown by decomposing $\epsilon=\epsilon_1+\epsilon_2$, where $\epsilon_1$ is a trial function of the form $\epsilon_1 = \epsilon_\infty +O(r^{-1})$, so that $\epsilon_2\sim O(r^{-1})$ for large $r$.  Then $\epsilon_2$ can be written formally as
\begin{align}\label{eq:eps2}
\epsilon_2(x) = \int dy \, G(x,y) [\slashed \nabla \epsilon_1](y),
\end{align}
were $G$ is a Green's function $G$ satisfying
\begin{align}
\slashed \nabla G(x,y) = \delta(x,y).
\end{align}
In other words, $G = D^{-1}$, where $D$ is the differential operator $\slashed \nabla$ restricted to spinors that vanish as $r\to \infty$.  Witten argues that $G$ exists because $D$ has no zero modes and the operator $\slashed \nabla$ is formally (anti) self-adjoint (though see \cite{Parker:1981uy}).  Any zero mode $\tilde \epsilon$ of $D$ must satisfy \eqref{eq:Ege0} with $\epsilon_\infty =0$.  Thus
\begin{align} \label{eq:epseq0}
\int_\Sigma |\nabla \tilde\epsilon|^2 = 0,
\end{align}
which implies $\tilde \epsilon = 0$.

Let us now consider the situation in the presence of the wall.  As noted in section \ref{sec:GD}, the energy $E$ then receives a contribution from the Brown-York stress tensor ${\cal T}_{AB}$  \cite{Brown:1992br} at the wall in addition to the familiar ADM-like boundary term $E_\infty$ at infinity.  We should thus impose a boundary condition at the wall such that this term appears in the analogue of
\eqref{eq:Ege0}.  As noted in \cite{Gibbons:1982jg}, this boundary condition should restrict half of the components of $\epsilon.$   We therefore impose the condition
\begin{align} \label{eq:epsDbc}
\left. P_- \epsilon\right|_{\partial\Sigma_0} = 0,
\end{align}
where $\partial\Sigma_0$ is the Dirichlet wall and
\begin{align}
P_- = \frac{1}{2} \left( 1 - \hat n_i \gamma^i\right) \, .
\end{align}
This choice was also used in \cite{Gibbons:1982jg}.  It leads to
\begin{align} \label{eq:epsQge0}
(\epsilon_\infty \epsilon_\infty^\dagger)E_\infty + \int_{\partial\Sigma_0} {\cal T}_{0 A} \bar \epsilon \gamma^A \epsilon  \ge \int_\Sigma |\nabla \epsilon|^2 \ge 0 ,
\end{align}
where the two terms on the left sum to the total energy $(\epsilon_\infty \epsilon_\infty^\dagger)E$ as desired.

One would then like to show that there exists a Green's function $G = D^{-1}$, where $D$ is now the differential operator $\slashed \nabla$ restricted to spinors that vanish as $r\to \infty$ and satisfy~\eqref{eq:epsDbc}.  But the $\partial \Sigma_0$ boundary term in \eqref{eq:epsQge0} is an obstacle, which in fact causes two problems.  First, note that it arises from integrations by parts.  This indicates that $\slashed \nabla$ generally fails to be formally (anti-) self-adjoint under the boundary condition \eqref{eq:epsDbc}.  This was not a problem in \cite{Gibbons:1982jg} where the analogue of $\partial \Sigma_0$ was a surface on which ${\cal T}_{AB} =0$.  The $\partial \Sigma_0$ boundary term also obstructs the proof that $\slashed \nabla$ has no zero modes, as instead of~\eqref{eq:epseq0} one finds only
\begin{align}
\int_{\partial\Sigma_0} {\cal T}_{0 A} \bar{\tilde \epsilon} \gamma^A \tilde\epsilon  \ge \int_\Sigma |\nabla \epsilon|^2 \ge 0\, .
\end{align}

It thus appears that this method yields no positivity result for spacetimes with Dirichlet walls.    Our original purpose in investigating this issue was to show that there is no tension between our negative kinetic energies and the $E \ge Q$ argument of \cite{Gibbons:1982fy} which holds in spacetimes with no such walls.  But the obstructions above also raise the question of whether the wall may in fact make the gravitational Hamiltonian unbounded below.

\section{Convergence of the derivative of $\psi$} \label{app:convergence}

In the main text we showed that the sequence $\{ \Delta\psi_k\}$ is uniformly convergent to zero as $k\to \infty$.  We now show that the same is true of derivatives $\grad^{(n)} \Delta\psi_k$ and we place a bound on the rate of convergence, i.e. we derive~\eqref{eq:psiBoundn}.

The proof of this convergence relies on the fact that $\grad_i \Delta \psi_k$ is also a solution to the Laplace equation (since $\grad$ commutes with the Laplacian) and the mean value property of harmonic functions.  Using these two facts, and letting $\Delta\psi_k = \psi -\psi_k$, it follows that
\begin{align}
\partial_i \Delta\psi_k(x) &=\frac{3}{4\pi r^3} \int_{B(x,r)} \partial_i \Delta\psi_k =\frac{3}{4\pi r^3} \int_{\partial B(x,r)} \hat r_i \Delta\psi_k ,
\end{align}
where $B(x,r)$ is a ball centered at $x$ with radius $r$ and $\hat r$ is unit normal to $\partial B$.  Taking the absolute value of both sides, and using $|\hat r_i| \le 1$ we obtain
\begin{align}
|\partial_i \Delta\psi_k(x)| & \le \frac{3}{r} \max_{\partial B}(|\Delta \psi_k|) < \alpha \left( \frac{3}{r}\right) \lambda^{(1-\delta)(k+1)},
\end{align}
for $k\ge 1$, where the last inequality follows from~\eqref{eq:psiBound}.  Iterating this procedure gives
\begin{align}
|\grad^{(n)} \Delta\psi_k(\vec x)|
\le \left(\frac{3}{r} \right) \max_{\partial B}(|\grad^{(n-1)} \Delta\psi_k(x)|) < \alpha \left( \frac{3}{r}\right)^n \lambda^{(1-\delta)(k+1)},
\end{align}

Now let $\vec y$ be an arbitrary point that is at least a distance $h/2$ away from the boundary $\partial \Sigma$.  Then, letting $r = h/2 $ we obtain
\begin{align} \label{eq:psiPartialBound}
|\grad^{(n)} \Delta\psi_k(\vec y)| < \alpha \left( \frac{6}{h}\right)^n \lambda^{(1-\delta)(k+1)}
\end{align}
We will now use the analyticity of harmonic functions and~\eqref{eq:psiPartialBound} to obtain bound on $|\grad^{(n)} \Delta\psi_k|$ over all of $\Sigma$.  Since $\grad^{(n)} \Delta \psi_k$ is a solution to the Laplace equation it is analytic in the domain of convergence of it's Taylor series, given schematically by
\begin{align}
\grad^{(n)} \Delta \psi_k(\vec y+\vec \epsilon\,) = \sum_{m=0}^\infty \frac{(\vec \epsilon \cdot \grad)^m \left(\grad^{(n)} \Delta\psi_k(\vec x)\right)}{m!}
\end{align}
Using~\eqref{eq:psiPartialBound} we obtain
\begin{align}
|\grad^{(n)} \Delta \psi_k(\vec y+\vec \epsilon\,)| < \frac{\alpha  \lambda^{(1-\delta)(k+1)}}{(h/6)^n}  \left[\sum_{m=0}^\infty \frac{1}{m!} \left(\frac{6|\vec \epsilon\, |}{h}\right)^m\right].
\end{align}
To cover all of $\Sigma$ we only need to consider $|\vec \epsilon \,| \le h/2$ which gives
\begin{align}
|\grad^{(n)} \Delta \psi_k(\vec y+\vec \epsilon\,)| < \frac{\alpha  \lambda^{(1-\delta)(k+1)}}{(h/6)^n}  \left[\sum_{m=0}^\infty \frac{3^m}{m!} \right] =\frac{\alpha e^3 \lambda^{(1-\delta)(k+1)}}{(h/6)^n} .
\end{align}
Letting $\beta := \alpha e^3$ we obtain~\eqref{eq:psiBoundn}.

\section{Constructing $\dot \psi$ and $\vec K$}\label{app:Kvec}

In this appendix we justify the use of $\vec K_1$, defined in~\eqref{eq:Kapprox}, as an approximation of $\vec K$.  Examining~\eqref{eq:Qsimple} we see that we are only interested in the quantities $\grad\cdot \vec K$ and $|\grad \times \vec K|$.  As noted in the main text we can explicitly verify that $\dot \psi_1 := -\grad \cdot \vec K_1$ satisfies the boundary conditions~\eqref{eq:psidotcutoffoft} to order $O(a)$.  So, it only remains to check that $|\grad \times \vec K_1|$ faithfully approximates $|\grad \times \vec K|$.

Without loss of generality we may write
\begin{align}
\vec K = \grad U + \grad\times \vec C,
\end{align}
where $\grad^2 U = - \dot\psi$.
Thus $\grad \cdot \vec K = -\dot \psi$, so all that remains is to construct a $\vec C$ such that $\grad^2 \vec K = 0$.

A simple calculation reveals that
\begin{align}
\grad \times \vec K &= - \grad^{\, 2}( C \hat \phi) + \grad (\grad \cdot C \hat \phi) =  D \hat \phi \cr
\grad^{\, 2} \vec K &= \grad \dot \psi - \grad \times(D\hat \phi),
\end{align}
where $D$ is defined by the relation $D\hat \phi = \grad^{\, 2}(C\hat \phi)$.  So, we must now find a $D$ such that $\grad^{\, 2} \vec K$ vanishes.  We will show that such a $ D$ is given by
\begin{align} \label{eq:Ddef}
\grad^2 ( D \hat \phi) = 0, \qquad  \left. \hat r \cdot(- \grad \dot \psi + \grad\times  D\, \hat \phi) \right|_{\partial\Sigma} = 0 ,
\end{align}
where $\hat r$ is the unit normal to $\partial \Sigma$.  To see that this is sufficient let $\vec L := \grad^2 \vec K$.  We then have
\begin{align}
\grad \cdot \vec L = - \grad^2 \dot \psi = 0 , \qquad \grad \times \vec L = -\grad^2  (D\, \hat \phi) + \grad (\grad\cdot  D \, \hat \phi) = 0,
\end{align}
where we have used $\grad \cdot \vec K = -\dot \psi$ and $\partial_\phi D = 0$.  Therefore, $\vec L$ is the gradient of a solution to the Laplace equation, i.e.
\begin{align}
\vec L = \grad V, \quad \grad^2 V = 0.
\end{align}
Furthermore, a complete set of boundary data for $V$ is given by
\begin{align}
\left. \hat r \cdot \grad V \right|_{\partial \Sigma} &= \left. \hat r \cdot \grad^2 \vec K\right|_{\partial \Sigma}  \cr
&= \left. \hat r \cdot (-\grad \dot \psi + \grad \times  D \, \hat \phi)\right|_{\partial \Sigma}  = 0 .
\end{align}
This boundary data is consistent with the solution $V=0$ and so by the usual arguments, this must be the unique solution.  Therefore
\begin{align}
\grad^2 \vec K = 0.
\end{align}
It is now straightforward to compute
\begin{align}
D_1 := (\grad \times \vec K_1) \cdot \hat \phi,
\end{align}
and we find that $D_1$ satisfies the boundary conditions associated with $D$ up to terms of order $O(a)$.  Therefore we obtain the result
\begin{align}
\grad \cdot (\vec K- \vec K_1) \sim a, \qquad |\grad \times (\vec K- \vec K_1) |\sim a.
\end{align}
It is straightforward to extend this construction to an arbitrary $\vec K_k$ for which the errors described above would be $O(a^k)$, however higher order terms do not contribute to our final result.

\bibliographystyle{kp}

\bibliography{ModuliSpacev11.bbl}

\end{document}